\documentstyle[editedvolume,numreferences,epsf]{crckapb}

\begin{opening}
\title{FRIEDEL OSCILLATIONS IN LUTTINGER LIQUIDS}

\author{REINHOLD EGGER}
\author{HERMANN GRABERT}

\institute{Fakult\"at f\"ur Physik, Albert-Ludwigs-Universit\"at\\ 
Hermann-Herder-Stra{\ss}e 3\\
D-79104 Freiburg, Germany}

\end{opening}

\begin{document}
\begin{abstract}
We study the density disturbance of a correlated 
one--di\-men\-sion\-al electron liquid in the presence 
of a scatterer or a barrier.
The $2k_F$--periodic density profile away from the barrier 
(Friedel oscillation) is computed for arbitrary electron--electron
interaction and arbitrary impurity strength. We find that in 
presence of correlations, the Friedel oscillation decays slower than
predicted by Fermi liquid theory.
In the case of  a spinless Luttinger liquid characterized by an interaction
constant $g\leq 1$, the asymptotic decay of the Friedel oscillation
is $x^{-g}$. For a weak scatterer, the decay is even slower
at small--to--intermediate distances from the impurity, with a crossover
to the asymptotic $x^{-g}$ law.
\end{abstract}
\section{INTRODUCTION}
One-dimensional (1D) interacting fermions have attracted a great deal
of interest in the recent past spurred by the 
observation that the bosonization technique developed  a long time 
ago \cite{luther}\nocite{mattis,emery}--\cite{haldane}
can be used to study the interplay between Coulomb interactions and 
disorder
\cite{kane92}\nocite{furu1,glaz2,furu2,fabrizio,sassetti,gog2}--\cite{maura}. 
Theoretical investigations of quantum wires are also
stimulated by the possibility of
performing experiments probing such systems \cite{jap}.
The last few years have seen a tremendous amount of activity,
mostly devoted to transport quantities like the 
conductance \cite{kane92}--\cite{leung}.
In this work, we focus on the equilibrium
electron {\em density}\, distribution of an interacting system with 
broken translational invariance \cite{prl}. 
In the presence of an impurity, conduction electrons will rearrange in order
to screen the impurity charge. For dilute impurity concentration,
it makes sense to first study the effect of a single impurity.
Depending on the impurity strength, a barrier
of arbitrary transmittance between zero and one 
hinders the electron flow. The corresponding crossover
between an insulator and a metal can significantly be affected by
the Coulomb interaction between electrons. This issue touches upon
important physical effects such as the
pinning of a Wigner crystal \cite{pin} or 
of charge density waves \cite{gruner}, the
breakdown of charging effects with increasing tunnel conductance \cite{sct},
or quantum transport in 1D heterostructure channels \cite{timp,gogolinrev}.

In the noninteracting case, the impurity is known 
to lead to  a $2k_F$--periodic
decaying disturbance of the ground state density, the so--called
Friedel oscillation \cite{friedel}. In $d$ dimensions, it has
far away from the barrier the asymptotic form 
\begin{equation} \label{friedel1}
\delta \rho (x) \sim \frac{\cos(2k_F x + \eta_F^{})}{x^d} \;,
\end{equation} 
where $\eta_F^{}$ is a phaseshift. In dimensions $d>1$, the Coulomb
interactions can presumably be incorporated by Fermi liquid theory
such that the asymptotic $x^{-d}$ behavior is still valid. 
Since Fermi liquid theory breaks down in 1D, the asymptotic decay law
of the Friedel oscillation can now be modified by interactions.
Indeed, as will be shown below, due to reduced 
screening, the asymptotic decay is always {\em slower} than the 
Fermi liquid $1/x$ law. 

In order to properly describe Coulomb interactions, one has to specify the
setup under consideration. If one deals with a 1D channel in gated
heterostructures (quantum wire) \cite{timp,gogolinrev}, the interactions
are usually screened due to the presence of metallic gates near the
channel. A short--ranged interaction leads to the 
Luttinger liquid model \cite{haldane} described by a single
dimensionless interaction constant $g$. The noninteracting Fermi liquid case
corresponds to $g=1$, and the presence of (repulsive) Coulomb interactions
implies $g<1$. On the other hand, one might as well consider a clean
isolated channel where the $1/r$ tail of the Coulomb potential is not
screened \cite{wigner}. We will refer to this situation as 
the ``long--ranged case'' in the following. 
For the sake of clarity, we shall only consider a single transport 
channel, and most of our analysis will be concerned with
the simplest case of spinless fermions.
Electron--electron backscattering effects can then be incorporated
by a renormalization of the interaction parameter $g$ \cite{solyom}.

Qualitatively, Coulomb interactions favor a smooth density profile. 
Thus one expects that the Friedel oscillation should be
smoothed and decay more slowly than predicted by Eq.~(\ref{friedel1}).
Furthermore, since the Coulomb interactions lead to a vanishing ground--state
conductance in the presence of an impurity \cite{kane92}, 
one should expect a more efficient pinning of the
Friedel oscillation, i.e., a larger amplitude than in the noninteracting
case. Our results corroborate this simple picture. For the case of a 
spinless Luttinger liquid,  the $1/x$ law is changed to an asymptotic
$x^{-g}$ decay far away from the barrier \cite{prl}.  For a weak 
scatterer, the behavior of the Friedel oscillation is quite complicated.
The decay is even slower than $x^{-g}$ for small--to--intermediate 
distance from the barrier, with a crossover to the asymptotic $x^{-g}$
law at some scale $x_0 \sim \lambda^{-1/(1-g)}$, where $\lambda$ denotes a
dimensionless impurity strength introduced below. 
These results have intrinsically many--body character
and cannot be found from Hartree--Fock or related techniques. 

In the presence of two impurities new features arise due to the
possibility of resonant tunneling
through the double barrier structure \cite{rt1}. For the noninteracting case,
it is easily shown that the Friedel oscillation on resonance decays
faster than $x^{-1}$, namely with a $x^{-2}$ law resulting from the
interference of both contributions.
Interactions will again modify the asymptotic properties
of the on--resonance Friedel oscillation.

\subsection{Model Hamiltonian}

We treat the 1D interacting electron liquid in the framework of 
standard bosonization \cite{luther}--\cite{haldane}.
This approach is appropriate for low temperatures, where only
excitations near the Fermi surface are relevant. In the following, 
we will mainly discuss the spinless case in detail. The creation
operator $\psi^\dagger(x)$ for spinless fermions can equivalently be
expressed in terms of the boson phase fields $\theta(x)$ and 
$\phi(x)$, which fulfill the algebra (we put $\hbar=1$)
\begin{eqnarray}\nonumber
[\theta(x) , \theta(x')]_- &=& [\phi(x) , \phi(x')]_- = 0  \\  \label{alg}
 \,[ \phi(x) , \theta(x')]_- & = & -(i/2) \, {\rm sgn} (x-x') \;.
\end{eqnarray}
Therefore $\Pi(x)=\partial_x\phi(x)$ is the canonically conjugate
 momentum to $\theta(x)$.
With $\psi^\dagger(x)=\psi^\dagger_+(x)+\psi^\dagger_-(x)$, where
the right-- and left--moving parts are given by
\begin{equation} \label{rlmovers}
\psi^\dagger_\pm(x) = \sqrt{\frac{\omega_c}{2\pi v_F}} \,
\exp[\pm i k_F x +  i\sqrt{\pi } (\phi(x)\pm\theta(x))]\;,
\end{equation}
the density operator $\rho(x)=\psi^\dagger(x) \psi(x)$ is readily found
in the form
\begin{equation} \label{bosondens}
\rho(x) = \frac{k_F}{\pi} + \frac{1}{\sqrt{\pi}} \partial_x \theta(x)
+ \frac{k_F}{\pi} \cos[2k_F x+ 2\sqrt{\pi} \theta(x)] \;.
\end{equation}
This boson representation of the electron density operator
 is of essential importance for our work. The three terms
are as follows. (1) The background charge density is $k_F/\pi$.
(2) The density due to right-- and left--movers results in the second term.
(3) The last term originates from the mixed terms, i.e., from the
interference between  right-- and left--movers. This $2k_F$--term is 
responsible for the Friedel oscillation, since it has a nonzero expectation
value if translational invariance is broken.
The bandwidth cutoff $\omega_c$ is defined as
\begin{equation}\label{wcdef}
\omega_c= v_F k_F \;,
\end{equation}
which is equal to the Fermi energy for the Tomonaga dispersion relation.

The effective low--energy
theory for a clean noninteracting fermion liquid is \cite{haldane}
\[
H_0 = \frac{v_F}{2} \int dx\, [ \Pi^2 + (\partial_x\theta)^2]  \;.
\]
Neglecting electron--electron backward--scattering processes,
the interaction among electrons
is then described by the density--density interaction term
\[
H_C = \frac{1}{2\pi} \int dx  dx' \, 
 \partial_x \theta(x) U(x-x')\partial_{x'}
\theta(x') \;,
\]
where $U(x-x')$ is the (screened) Coulomb interaction potential. Most of 
our analysis is concerned with the Luttinger liquid case, where 
one has short--ranged interactions.  We are then led to the  
Luttinger liquid Hamiltonian
\[
H_L = \frac{v_F}{2} \int dx \left[  \Pi^2 + \frac{1}{g^2} 
(\partial_x\theta)^2\right]  \;,
\]
where the interaction parameter
\[
  g = \frac{1}{\sqrt{1+ U_0/\pi v_F}} \leq 1 
\]
is related to the forward scattering amplitude $U_0$,
such that $g=1$ represents the noninteracting case. 

Let us now consider an elastic potential scatterer at $x=0$ described
by a potential $V(x)$. It leads to a 
contribution $H_I = \int dx V(x) \rho(x)$,
and for a single $\delta$-scatterer, $V(x)= \pi V k_F^{-1} \delta(x)$,
one finds from Eq.~(\ref{bosondens}) the generic form
\begin{equation} \label{impurity}
H_I = \frac{\sqrt{\pi}\, V}{k_F} \partial_x \theta(0)+
 V \cos[2\sqrt{\pi}\theta(0)]\;.
\end{equation}
The impurity strength $V$ will often be given as dimensionless quantity
\begin{equation} \label{imps}
\lambda= \pi V/\omega_c\;.
\end{equation}
Tuning $\lambda$ from zero to infinity corresponds to changing the
transmittance of the barrier from unity down to zero. 
The Hamiltonian is then $H=H_0+H_C+H_I$, i.e., 
\begin{eqnarray}\label{boson}
H &=& \frac{v_F}{2} \int dx\, [ \Pi^2 + (\partial_x\theta)^2] 
 + \frac{1}{2\pi} \int dx  dx' \,  
 \partial_x \theta(x) U(x-x')\partial_{x'}\theta(x') \nonumber
\\ &+& \frac{\sqrt{\pi}\, V}{k_F} \partial_x \theta(0) + 
V \cos[2\sqrt{\pi}\theta(0) ] \;.
\end{eqnarray}
In the Luttinger liquid case the first two terms are replaced by
$H_L$. The model (\ref{boson})
has been the subject of many studies in the past few years, primarily 
with regard to conductance 
computations \cite{kane92}\nocite{furu1,glaz2,furu2}--\cite{fabrizio}.

\subsection{Generating Functional}

As is apparent from Eq.~(\ref{bosondens}), one can compute
 $\langle\rho(x=y)\rangle $ from the generating functional
$Z(y) = \langle \exp [2\sqrt{\pi} \,i \mu \theta(y)] \rangle$.
We can first gauge away the forward-scattering term $\sim \partial_x
\theta(0)$ in $H_I$ by the unitary transformation $U=\exp[ig^2 \lambda
\phi(0)/\sqrt{\pi}]$. Then $Z(y)$ becomes
\begin{equation}\label{gener}
Z(y) = \langle \exp [2\sqrt{\pi} \,i \mu \theta(y)] \rangle
\,e^{-i\mu g^2 \lambda \,{\rm sgn}(y)} \;,
\end{equation}
where the average has to be carried out using $UH U^{-1}$, which
is just Eq.~(\ref{boson}) without the forward-scattering term.

We formally solve for $Z$ by introducing a field $q(\tau)$, which is
constrained by
\[
q(\tau)= 2\sqrt{\pi}\, \theta(x=0,\tau) \;.
\]
This constraint is enforced by a Lagrange multiplier field $\Lambda(\tau)$,
such that one has the effective Euclidean action
\begin{eqnarray}
\nonumber
S_e[\theta,\Lambda,q]
&=& \frac{v_F}{2} \int dx d\tau \left [ \frac{1}{v_F^2} (\partial_\tau
\theta )^2 + (\partial_x\theta)^2 \right] \nonumber
\\ &+& \frac{1}{2\pi} \int dx  dx' d\tau \,  
 \partial_x \theta(x,\tau) U(x-x')\partial_{x'}\theta(x',\tau) \nonumber
\\&+&  V \int d\tau \, \cos q(\tau) \nonumber
\\ &-&2\sqrt{\pi}\, i \mu \theta(y,0)
+  i \int d\tau\, \Lambda(\tau)\,  [2\sqrt{\pi}
\,\theta(0,\tau) - q(\tau)]  \;. 
 \label{seff}
\end{eqnarray}
The $\theta$ part of this effective action is Gaussian and can therefore
be treated exactly by solving the classical Euler--Lagrange equation,
\begin{equation} \label{euler}
\frac{1}{v_F^2} \frac{\partial^2 \theta}{\partial \tau^2}  +
\frac{1}{g^2}\frac{\partial^2 \theta}{\partial x^2}
= \frac{2\sqrt{\pi} i}{v_F} \, [ \delta(x) \Lambda(\tau) -
 \mu \delta(x-y) \delta(\tau) ] \;.
\end{equation}
The solution of Eq.~(\ref{euler}) is easily found in Fourier space,
\[
\theta(x,\tau) = \int \frac{d\omega}{2\pi}\int \frac{dk}{2\pi}
e^{i\omega \tau + ikx} \,\theta (k,\omega)\;,
\]
and similarly for $\Lambda(\tau)$. Then Eq. (\ref{euler}) takes the form
\[
(\omega^2 + \omega_k^2) \, \theta(k,\omega) = 
-2\sqrt{\pi} \,i\,v_F\, (\Lambda(\omega) - \mu \,e^{-iky} ) \;,
\]
where we have introduced the plasmon frequency 
\begin{equation} \label{plasmon}
\omega_k = v_F |k| \sqrt{1+U_k/\pi v_F} 
\end{equation}
with $U_k=\int dx \exp(-ikx) U(x)$. In the Luttinger liquid case,
this is simply $v_s |k|$ with the sound velocity $v_s=v_F/g$.
Defining the boson propagator functions
\begin{eqnarray} \label{bosprop}
F(x,\omega) &=& v_F \int_{-\infty}^\infty dk\, \frac{\cos(kx)}{\omega^2 + 
\omega_k^2} \\ \label{bosll}
&=& \frac{\pi g}{|\omega|} \, e^{-|g \omega x|/v_F} \qquad 
{\rm (Luttinger \; liquid)}\;,
\end{eqnarray}
one finds the solution 
\begin{equation} \label{clasol}
\theta(x,\tau) = \frac{-i}{\sqrt{\pi}} \int
 \frac{d\omega}{2\pi}\, e^{i\omega \tau} 
\, [ \Lambda(\omega) F(x,\omega) - \mu F(x-y,\omega) ].
\end{equation}
Next we have to determine the action corresponding to the
classical solution. Inserting  Eq.~(\ref{clasol}) into 
Eq.~(\ref{seff}), we find after some algebra
\begin{eqnarray*}
S_{\rm cl}[\Lambda,q] &=& \int \frac{d\omega}{2\pi} 
\Bigl\{ [\mu^2 + \Lambda(\omega)\Lambda(-\omega)] F(0,\omega) 
- 2 \mu \Lambda(-\omega) F(y,\omega)\\  &-&i \Lambda(-\omega)
q(\omega) \Bigr\}
+ V \int d\tau \,\cos[q_0 + q(\tau)] - i\mu q_0  \;,
\end{eqnarray*}
where $q_0$ is the zero--mode of the auxiliary field.
Since $S_{\rm cl}$ is quadratic in $\Lambda$, 
the Lagrange multipliers are simply found by extremization.
The result is 
\[
\Lambda(\omega) = \mu \frac{F(y,\omega)}{F(0,\omega)} +
i \frac{q(\omega)}{2 F(0,\omega)}\;,
\]
and inserting this into $S_{\rm cl}$ gives the
generating functional (\ref{gener}) in the form of an average over the
$q$ field.

In the end, the generating functional takes the form 
\begin{equation}  \nonumber
Z(x) = W(x)^{\mu^2} e^{-i\mu g^2\lambda \,{\rm sgn}(y)} \left \langle
\exp\left[i\mu \left( q_0 + \int \frac{d\omega}{2\pi}
q(\omega) \frac{F(x,\omega)}{F(0,\omega)}\right)\right] \right\rangle_q \;,
\label{gen}
\end{equation}
where the  remaining $q$ average has to be taken with the action
\[
S[q]= \int \frac{d\omega}{2\pi} \frac{q(\omega) q(-\omega)}
{4 F(0,\omega)} + V \int d\tau \, \cos[q_0  +q(\tau) ] \;.
\]
For a Luttinger liquid, we obtain therefore
\begin{equation} \label{dissac2}
S[q]= \int_{-\infty}^\infty \frac{d\omega}{2\pi} \frac{|\omega|}{4\pi g} \, 
 |q(\omega)|^2  +     V \int d\tau \, \cos[q_0 +q(\tau) ] \;.
\end{equation}
Apart from the kinetic energy term, this action corresponds to the action of
a Brownian particle ($q$ translates into the position of the particle)
moving in a cosine potential under the influence of Ohmic 
dissipation \cite{guinea}\nocite{schmid,fisher}--\cite{weiss}.
The damping strength is connected with the Coulomb interaction constant
$g$, and the height of the cosine potential is related to the impurity
strength $\lambda$.
The finite mass of the particle can be associated with the bandwidth 
$\omega_c$.

The {\em envelope function} in Eq.~(\ref{gen})
\begin{equation} \label{envel}
W(x) = \exp\left [ \int\frac{d\omega}{2\pi}
 \frac{F^2(x,\omega) - F^2(0,\omega)}
{F(0,\omega)}  \right]
\end{equation}
does not depend on impurity properties at all and involves no $q$ averaging.
It will turn out that this function governs the asymptotic properties of the
Friedel oscillation. 

\section{FRIEDEL OSCILLATIONS} 

In this section, we discuss the ground--state equilibrium 
Friedel oscillation induced by a barrier of arbitrary strength 
$\lambda=\pi V/\omega_c$ in the presence of Coulomb interactions.
Most of our analysis will focus on the case of a spinless Luttinger liquid,
but we mention some generalizations to the spin--$\frac12$ situation or
to long--ranged interactions. Since the density profile
is symmetric around $x=0$, we put $x\geq 0$ in the following.

\subsection{General expression}

In the case of a spinless Luttinger liquid characterized by the interaction
constant $g$, the boson propagator 
takes the form given in Eq.~(\ref{bosll}), namely
\[
F(x,\omega) = \frac{\pi g}{|\omega|} \exp[-|\omega|x/v_s]\;.
\]
The envelope function (\ref{envel}) is therefore given by
\[
W(x) = \exp\left[ g \int_0^\infty d\omega \frac{ 
e^{-2\omega x/v_s} -1}{\omega}  \right] \;,
\]
and we have to introduce an ultraviolet cutoff to regularize the 
integral. The appropriate cutoff $\omega_c$ is provided by the Fermi energy
[see Eq.~(\ref{wcdef})], and employing an exponential cutoff function
$\exp[-\omega/\omega_c]$, we obtain
\[
W(x) = (1 + x/\alpha)^{-g} \;,
\]
with the microscopic lengthscale $\alpha = v_F/2g\omega_c$.
Defining a
``lattice spacing'' $a=\pi v_F/\omega_c$, one has $\alpha=a/2\pi g$.

Let us now compute the deviation in the electron density profile,
 $\delta \rho(x) = \langle\rho(x)\rangle - k_F/\pi$,
caused by the presence of the impurity. From Eq.~(\ref{gen}), one 
can verify that away from the impurity 
the slow $(k\approx 0)$ component in Eq.~(\ref{bosondens})
is not affected by a potential scatterer. 
The only space--dependent density profile response to the 
impurity is the Friedel oscillation,
\begin{equation}
\delta \rho(x) /\rho_0 = - 
  (1+x/\alpha)^{-g} \, \cos(2k_F x - g^2 \lambda)\, P(x) \;,
\end{equation}
where $\rho_0=k_F/\pi$.
There is a renormalization of the
noninteracting phase shift $\eta_F=\lambda$,
which becomes $g^2 \lambda$ in the case of a Luttinger liquid.

 The {\em pinning function} $P(x)$
includes the nontrivial $q$ average and takes the form
\begin{equation} \label{pinning}
P(x) = - \left \langle \cos\left[q_0+\int \frac{d\omega}{2\pi}
 e^{-g|\omega| x/v_F } q(\omega) \right] \right\rangle_q\;.
\end{equation}
A useful quantity is the {\em pinning amplitude}
\[
P_0 = -\delta\rho(x=0)/\rho_0 = P(0) \;,
\]
and it is also convenient to define the quantity 
\[
P_\infty = P(x\to \infty) = - \langle \cos q_0 \rangle_q\;.
\]
From the definition of the pinning function, it is clear that
$0\leq P(x)\leq 1$ must be always fulfilled. Furthermore,
the pinning function increases monotonically from $P_0$ to
$P_\infty$.

For transmittance one ($\lambda=0$), the ``charge'' $q$ is free and $P(x)=0$.
For zero transmittance ($\lambda\to \infty$), 
the potential $V \cos q$ locks $q$ at odd multiples of $\pi$, and $P$
takes its maximal value, $P(x)=1$, for all $x$.  In that case, we obtain
readily that the Friedel oscillation decays as $x^{-g}$. This result
for zero transmittance can also be obtained by open boundary 
bosonization \cite{gog2}. The slow algebraic decay
of the Friedel oscillation cannot be reproduced by
Hartree--Fock type calculations and is a true many--body effect.
For instance, this can explicitly be seen by considering 
the limit $\lambda\to \infty$,
where the Hartree--Fock type procedure devised by 
Matveev {\em et al.} \cite{matveev1} would predict a $x^{-1}$ decay.
In the following, we shall discuss the properties of the pinning 
function in some detail. 

\subsection{Noninteracting case: Fermionization}

Let us first discuss the exact solution of the bosonized model
(\ref{boson}) for the noninteracting case $g=1$. For this
particular value, we can obtain the exact solution for all
quantities of interest by means of fermionization.
This is seen by re--writing the Hamiltonian (\ref{boson})
for $g=1$ in terms of the right-- and left--moving fermion
operators $\psi^\dagger_\pm(x)$  as given in Eq.~(\ref{rlmovers}).
The impurity term is a product of fermion operators
at $x=0$, and the bulk term becomes the massless Dirac
Hamiltonian. In the end, $H$ is equivalently expressed
in the fermionized form
\[
H^f = -i v_F \int dx \sum_{p=\pm} p
\psi^\dagger_p(x) \frac{\partial}{\partial x}
\psi^{}_p(x)  + v_F \lambda \sum_{p,p'=\pm} \psi_p^\dagger(0) \psi_{p'}^{}(0)\;,
\]
where $\lambda=\pi V/\omega_c$ is the impurity strength (\ref{imps}).

Since $H^f$ is quadratic in the fermion operators, it is sufficient
to study the equations of motion
\begin{equation} \label{eqm}
\left ( \frac{1}{v_F} \frac{\partial}{\partial t} \mp
\frac{\partial}{\partial x}  \right) \psi^\dagger_\pm (x,t) =
i\lambda \delta(x) ( \psi^\dagger_+(x,t) + \psi_-^\dagger(x,t) ) \;.
\end{equation}
Away from $x=0$, the solutions of Eq.~(\ref{eqm}) are simply free waves
$\sim \exp(-iv_F k t \pm i kx )$, where $k>0$ is the wavevector.
At $x=0$, the right-- and left--moving components are discontinuous.
Adding the two equations (\ref{eqm}) and integrating over an
infinitesimal region around $x=0$, we obtain the jump condition
\[
\left( \psi_+^\dagger - \psi^\dagger_-\right)(0^+) -
\left( \psi_+^\dagger - \psi^\dagger_-\right)(0^-)
= -2 i \lambda  
\left( \psi_+^\dagger + \psi^\dagger_-\right) (0) \;,
\]
which implies that the $\psi_\pm^\dagger$ are given in terms of the
usual scattering waves \cite{messiah} with 
$k$--independent transmission and reflection amplitudes,
or equivalently a $k$--independent phase shift
\begin{equation}\label{trans2}
\eta_k = \eta_F^{} = \lambda \;.
\end{equation}
In the absence of correlations, all effects of the 
impurity are contained in this phase shift, and 
the Friedel oscillation is readily evaluated in closed form. 
By expressing the reflection amplitude $r_k$ in terms of the
phase shift $\eta_k$, and then using the relation \cite{messiah}
\[
\delta \rho (x) = \frac{{\rm Re}}{\pi} 
 \int_0^{k_F} dk\, r_k e^{2ik|x|}\;,
\]
we obtain the ground-state result
\begin{equation} \label{frg1}
\delta \rho(x) = \frac{\sin \eta_F^{}}{2\pi |x|}
\;[ \cos(2k_F|x|+\eta_F^{}) - \cos(\eta_F^{})  ] \;.
\end{equation}
As expected for the noninteracting case, the 1D Friedel oscillation
indeed decays as $1/x$.

\subsection{Self--consistent harmonic approximation (SCHA)}
\label{sec:scha}

Let us begin our discussion of the pinning function 
for $g<1$ by describing 
a simple approximation based on Feynman's variational principle
(self--consistent harmonic approximation, SCHA) \cite{fisher,gog1}.
The most important approximation made in the SCHA is the neglect
of tunneling transitions between different wells of the impurity cosine
potential which seems reasonable for large $\lambda$. 
Therefore, we can assume $q_0$ to be an odd multiple of $\pi$
and consider a Gaussian trial action
\begin{equation} \label{dissac3}
S_{\rm tr}[q]= \int_{-\infty}^\infty 
\frac{d\omega}{2\pi} \frac{|\omega|}{4\pi g} \, 
 |q(\omega)|^2  +     \frac{\Omega}{2} \int d\tau \, q^2(\tau) \;,
\end{equation}
where the frequency $\Omega$ is determined from a variational
principle for the free energy \cite{feynman}.
 It states that the free energy $F$
obeys the inequality
\[
F \leq F_{\rm tr} + \langle H - H_{\rm tr} \rangle_{\rm tr} \;,
\]
where the average has to be carried out using the trial action 
(\ref{dissac3}).
Minimization of $F$  yields 
\[
\Omega= V \exp\left [-\langle q^2 \rangle_{\rm tr} /2\right]\;.
\]
From  Eq.~(\ref{dissac3}), we can read off
\begin{equation} \label{av}
\langle q(\omega) q(-\omega') \rangle_{\rm tr}
= \frac{2\pi \delta(\omega'-\omega)}{\Omega + |\omega|/2\pi g}\;,
\end{equation}
such that 
\[
\langle q^2 \rangle_{\rm tr} = 2g \ln\left
 ( 1+ \frac{\omega_c}{2\pi g \Omega}\right)\;,
\]
and a self--consistency relation follows,
\begin{equation} \label{selfcon}
\Omega/V =  \left(1+\frac{\omega_c}{2\pi g \Omega}\right)^{-g}\\
= \left \{
\begin{array}{c@{\quad,\quad}l}
   1  & \lambda\gg 1 \\
  (2 g \lambda)^{g/(1-g)} & \lambda \ll 1 \;.
\end{array} \right. 
\end{equation} 
 For a strong scatterer $\lambda=\pi V/\omega_c
\gg 1$, the trial frequency
is simply $V$, as follows by a direct expansion of the impurity 
cosine term around the minima $q_0$, which are odd multiples of $\pi$.
In this limit, only small fluctuations about these minima are possible,
with interwell transitions being forbidden by an exponentially small
WKB tunneling factor. Therefore, we expect SCHA to be most valuable
for the strong--scattering limit.

From Eq.~(\ref{pinning}) together with Eq.~(\ref{av}), one can easily 
evaluate the now Gaussian average. SCHA yields 
for the pinning function,
\begin{equation} \label{respin}
P(x) = \exp\left[
 -g\, e^{(x+\alpha)/x_0}\, E_1((x+\alpha)/x_0) \right]  \;,
\end{equation}
with the exponential integral $E_1(y)$ \cite{abram}
and the {\em  crossover scale} $x_0$ given by
\begin{equation} \label{cross}
x_0/\alpha = \frac{1}{ 2  g\lambda} \;(V/\Omega)\\
= \left \{ \begin{array}{c@{\quad,\quad}l}
   1/(2g\lambda) & \lambda \gg 1 \\
   (2 g\lambda)^{-1/(1-g)} & \lambda \ll 1 \;,
\end{array} \right. 
\end{equation}
where we have used Eq.~(\ref{selfcon}) in the second step.

Since in the strong--scattering limit $x_0$ is even smaller than $\alpha$, 
the term ``crossover'' is not meaningful in this limit.
Using asymptotic properties of  $E_1(y)$, Eq.~(\ref{respin}) becomes 
for $x\gg \max(\alpha,x_0)$ 
\begin{equation} \label{Plarge}
P = e^{-g x_0/x} \simeq 1\;. 
\end{equation}
In the strong--scattering limit, the pinning function is essentially unity 
for all $x$, and the $x^{-g}$ decay is always found. This result
is in accordance with Monte Carlo results.

In the {\em weak--scattering limit}, $\lambda \ll 1$,  
the pinning function exhibits far more structure. 
The crossover scale goes to infinity as $\lambda\to 0$, namely
$x_0 \sim \lambda^{-1/(1-g)}$. One expects that there are 
two different types of behavior for $x\ll x_0$ and
$x\gg x_0$. Unfortunately, as  we will see below, 
SCHA is unable to provide correct quantitative results except
for very strong interactions, $g\ll 1$.  That SCHA becomes more accurate
for stronger interactions can be rationalized as follows. The presence
of interactions leads, loosely speaking, to a renormalization of the
barrier height. Using a perturbative renormalization
group (RG) approach \cite{kane92}, one finds that $\lambda$ grows under
the RG transformation.
For strong interactions, it flows quickly into the strong--scattering 
limit, where SCHA is essentially exact.

For $x\gg x_0$, SCHA always gives $P\simeq 1$
according to Eq.~(\ref{Plarge}). This is an incorrect result,
as can be seen from the exact result for the special case $g=1$.
That failure is due to the complete neglect of interwell tunneling
in the SCHA. Without tunneling transitions, $q_0$ must be an odd
multiple of $\pi$, and one finds $P_\infty=1$
as predicted by SCHA. However, taking into account 
excursions to neighboring wells, it is readily seen that in general
$P_\infty < 1$. 

Despite of these shortcomings, the effective Gaussian treatment 
indicates that for a weak scatterer, there is a crossover, 
with a {\em slower} decay of the Friedel
oscillation at small--to--intermediate distances $x\ll x_0$
than the asymptotic $x^{-g}$ decay. In fact, Eq.~(\ref{respin}) gives
\[
P (x) = [ (x+\alpha)/x_0]^g\;, \qquad x\ll x_0\;,
\] 
which would imply that the Friedel oscillation does {\em not decay}
at all up to the scale $x_0$. Only for $x\gg x_0$, one would
have the $x^{-g}$ law. 

In conclusion, for a weak scatterer, SCHA yields for 
the Friedel oscillation 
\begin{equation} \label{scha}
\delta\rho(x)/\rho_0 = - \cos(2k_F x - g^2 \lambda) 
\times \left \{ \begin{array}{c@{\quad,\quad}l}
   (x_0/\alpha)^{-g} & \alpha \ll x \ll x_0 \\ 
   (x/\alpha)^{-g} &  x\gg x_0 \;. \end{array} \right. 
\end{equation}
This prediction is certainly
incorrect for weak Coulomb interactions, see Eq.~(\ref{frg1}).
The effective Gaussian treatment is only 
valuable for a strong scatterer or for strong Coulomb
interactions. Furthermore,
SCHA provides an estimate for the important crossover scale $x_0$.

\subsection{Perturbation series for the pinning function}

To investigate the weak--scattering limit,
we next attempt to evaluate the pinning function by perturbation
theory in $\lambda$.
The perturbation series is found
 by expanding the impurity propagator \cite{schmid},
\[
\exp\left[-V\int d\tau\; \cos q(\tau) \right] = 
\sum_{m=0}^\infty 
(-V/2)^m  \int {\cal D}_m\tau \sum_{\{\sigma\}}
\exp\left[i\sum_{j=1}^m \, \sigma_j q(\tau_j) \right]\;,
\]
where one has to sum over all auxiliary variables $\sigma_j=\pm 1$,
and $\int {\cal D}_m\tau$ denotes a time--ordered integration over the $m$
possible intermediate times $\tau_j$. Thereby, the $q$ average becomes
Gaussian again, and $P(x)$ takes
 indeed the form of a power series in $\lambda$. One finds easily
that only odd powers in $\lambda$ contribute to the perturbational
expansion.

We find for the lowest--order contribution 
\begin{equation} \label{pert}
P(x) =\lambda  \gamma^{(1)}_g 
 \left(x/\alpha\right)^{1-g} + {\cal O}(\lambda^3)   \;,
\end{equation}
with the numerical prefactor
\begin{equation} \label{ggg}
\gamma^{(1)}_g =  \frac{4^{g-1}\,\Gamma(g-\frac12)}{\sqrt{\pi}\,\Gamma(g)}\;,
\end{equation}
where $\Gamma(z)$ is the Gamma function.
This perturbative result is only valid for $g>\frac12$, since 
otherwise $\gamma^{(1)}_g$ is not defined. That the computation of $P(x)$
is indeed a nonperturbative problem for all $g<1$ can be seen 
by computing the higher--order terms in $\lambda$.
From dimensional scaling, the perturbation series must have the form
\[
P(x) = \sum_{m=1}^\infty \lambda^{2m-1} 
\gamma^{(m)}_g \left({x/\alpha}\right)^{(2m-1)(1-g)}\;.
\]
Since the higher--order terms increase faster, 
the first--order estimate (\ref{pert})
can only be valid for $x\ll x_0$. The crossover scale $x_0$ may
be computed by equating the $m=0$ and $m=1$ components, or by 
arguing that $P(x)\leq 1$.
Apart from numerical prefactors of order unity,
these estimates for the crossover scale coincide 
with the SCHA crossover scale (\ref{cross}). 

In contrast to Eq.~(\ref{scha}), perturbation theory would 
therefore predict for the Friedel oscillation
in the case of a weak scatterer $\lambda\ll 1$,
\begin{equation} \label{pertu}
\delta\rho(x)/\rho_0 = - \cos(2k_F x - g^2 \lambda) 
\times\left \{ \begin{array}{c@{\quad,\quad}l}
 \lambda \gamma_g^{(1)} (x/\alpha)^{1-2g} & \alpha \ll x \ll x_0
 \\ P_\infty (x/\alpha)^{-g} &  x\gg x_0 \;.\end{array} \right. 
\end{equation}
Here, we have assumed that the pinning function
approaches a constant value $P_\infty \leq 1$
for $x\gg x_0$. This will be confirmed below by Monte
Carlo simulations.
The perturbative result (\ref{pertu}) is expected to hold at least
 for weak interactions, $1-g \ll 1$. Perturbation theory breaks
down even for small
 distances, $x\ll x_0$, in the case of strong interactions,
$g\leq \frac12$. 

In conclusion, perturbation theory predicts a similar 
weak--scattering scenario as SCHA. The Friedel oscillation
exhibits a slower decay $\sim x^{1-2g}$ at small--to--intermediate
distances from the barrier, with a crossover to the asymptotic $x^{-g}$ law.
The estimate for the crossover scale $x_0$ coincides in both approximations
up to a numerical constant of order unity.  From the scaling
$x_0 \sim \lambda^{-1/(1-g)}$,
one observes that the limit of a vanishing barrier implies a nontrivial
long--distance behavior of the pinning function.

\subsection{Monte Carlo results}
\label{sec:mc}

One can compute the pinning function (\ref{pinning})
for any barrier height $\lambda$
and arbitrary interaction constant $g$ by employing numerically
exact quantum Monte Carlo simulations.
As our results for the strong--scattering
limit corroborate the SCHA prediction $P(x)\simeq 1$, we 
only present numerical data for the weak--scattering case here.
Numerical simulations are particularly useful 
for a determination of the behavior of the pinning function
at small--to--intermediate distances
from the impurity, $x\ll x_0$, where we expect to find a power law
\begin{equation}\label{exponent}
P(x) \sim x^{\delta_g}\;.
\end{equation}
The two approaches discussed above suggest a 
$\lambda$--independent exponent
$\delta_g = g$ (SCHA) and $\delta_g = 1-g$ (perturbation theory).
The SCHA estimate should be valid for strong interactions, $g\ll 1$,
while perturbation theory should hold at least for weak interactions,
$1-g\ll 1$. 

\begin{figure}
\epsfysize=8cm
\epsffile{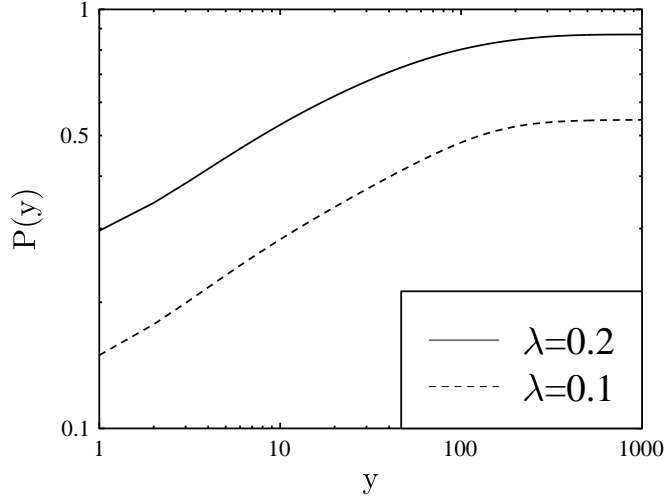}
\caption[]{\label{fig1} Monte Carlo data for the pinning function $P(y)$
at $g=\frac12$ for two different values of the  impurity strength. The 
dimensionless space variable  is $y=\omega_c x/v_F$. Notice the logarithmic
scales.}
\end{figure}

\begin{figure}
\epsfysize=8cm
\epsffile{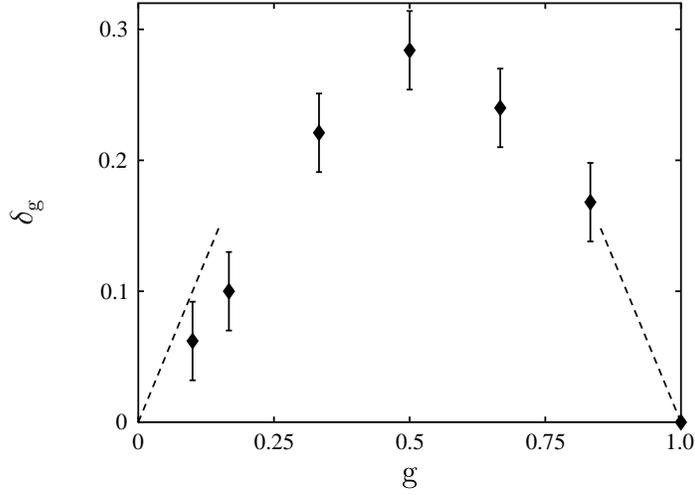}
\caption[]{\label{fig2} Numerical estimates for the exponent $\delta_g$.
Error bars were obtained from comparing data for different $\lambda$.
The dashed curves are the SCHA estimate $\delta_g=g$ and the 
perturbative estimate $\delta_g=1-g$, respectively. }
\end{figure}

To compute the pinning function by Monte Carlo, one has to consider
finite temperatures, such that the frequency integrals become sums over
Matsubara frequencies, $\omega_n=2\pi n/\beta$, with $\beta=1/k_B T$.
We have checked that the temperatures used in our simulations were
low enough to ensure that one is in the zero--temperature limit. Typically,
$\beta \omega_c= 1000$ was sufficient. We have employed a hard cutoff
scheme by keeping only the Matsubara frequencies with $|\omega_n|<
\omega_c$. The Matsubara components $q(\omega_n)$ are then sampled
according to the action (\ref{dissac2}) using the standard Metropolis 
algorithm. 
Our data were obtained on an IBM RISC
6000/Model 590 workstation using 50.000 samples for a given parameter set,
with subsequent samples separated by 5 passes.  One
can compute the full pinning function for all $x$ of interest
in a single simulation run.

In Fig.~\ref{fig1}, we show data for $g=\frac12$ and the two 
impurity strengths $\lambda=0.1$ and $\lambda=0.2$.
The data clearly display the
crossover. We find the power law (\ref{exponent})
for $x\ll x_0$, with an exponent
$\delta_{1/2} = 0.28\pm 0.03$. For $x\gg x_0$, the pinning function
is essentially constant. 
The asymptotic decay of the Friedel oscillation at large 
$x\gg x_0$ is therefore
$x^{-g}$, while the decay is slower for $x\ll x_0$,
\[
\delta \rho(x) \sim x^{-g+\delta_g}\;.
\]
As is apparent from Fig.~\ref{fig1}, 
the region where this law is valid shrinks 
rapidly as the impurity strength is increased. Our data are generally
consistent with the scaling $x_0\sim \lambda^{-1/(1-g)}$.

\begin{figure}
\epsfysize=8cm
\epsffile{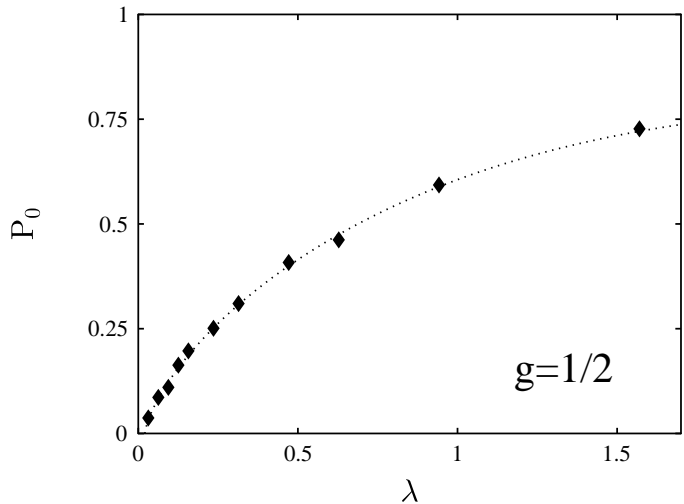}
\caption[]{\label{fig3} Monte Carlo data for the 
pinning amplitude $P_0$ at $g=\frac12$ as a 
function of the impurity strength $\lambda$. The dotted curve
is a guide to the eye only.}
\end{figure}

Fig.~\ref{fig2} displays numerical values for $\delta_g$. From our data,  the 
power--law exponent $\delta_g$ appears to be independent 
of $\lambda$.   The SCHA prediction $\delta_g=g$
is valid asymptotically for $g \rightarrow 0$, while for $1-g\ll 1$, 
the perturbative result $\delta_g=1-g$ is reproduced, with 
a smooth turnover between both limits. 

\begin{figure}
\epsfysize=8cm
\epsffile{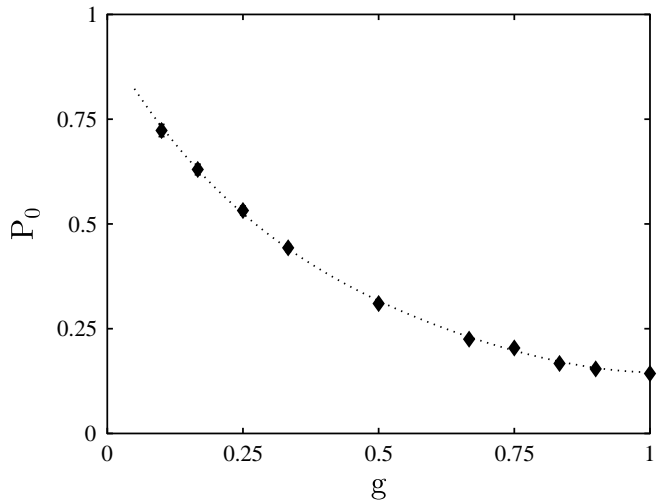}
\caption[]{\label{fig4} Monte Carlo data for the pinning amplitude  as 
a function of the interaction constant $g$ at fixed
impurity strength $\lambda=\pi/10$. The dotted curve is a guide to the eye
only.}
\end{figure}

\begin{figure}
\epsfysize=8cm
\epsffile{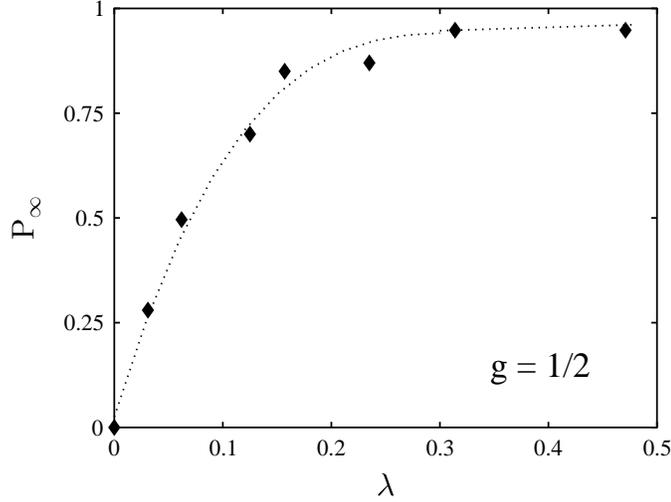}
\caption[]{\label{fig5} Monte Carlo data
 for $P_\infty$ at $g=\frac12$ as a function
of the impurity strength. The dotted curve is a guide to the eye only.}
\end{figure}

The ability of the scatterer to pin the charge density wave can 
conveniently be measured in terms of the pinning amplitude $P_0$.
As expected, $P_0$ increases with increasing impurity strength  $\lambda$,
 see Fig.~\ref{fig3}. To study the dependence of $P_0$ on the 
interaction strength $g$, we show in Fig.~\ref{fig4} data for $P_0$ at fixed impurity 
strength. Clearly, the pinning is more efficient as
the interaction becomes stronger.
As mentioned in the introduction, this can qualitatively be understood
in terms of the scaling of the barrier height \cite{kane92}.
Finally, in Fig.~\ref{fig5}, we show the quantity $P_\infty$ for 
$g=\frac12$ as a function of $\lambda$. It shows
the same qualitative behavior as in the noninteracting case. 
Most importantly, we generally find $P_\infty< 1$, in 
contrast to the SCHA prediction. Deviations from unity directly
reflect tunneling transitions between different wells of the 
cosine potential.  

One might wonder about  computing the pinning function exactly at the
special value $g=\frac12$. Unfortunately, the  fermionization 
method which is able to yield exact results for the conductance \cite{guinea}
cannot achieve this aim (except for the quantity $P_0$), since
the density operator (\ref{bosondens}) cannot be expressed in terms
of the fermions employed to diagonalize the $g=\frac12$ model. In contrast,
for vanishing interaction at the point $g=1$, the fermionization method 
allows for an exact computation of the pinning function. To substantiate
our results it would be of much interest to apply perturbed conformal field 
theory methods together with the thermodynamic Bethe ansatz
to the interacting problem.

\subsection{Long--ranged interaction}
\label{sec:lr}

In the absence of metallic gates, one has to take into account
the long--ranged character of the Coulomb interaction
between the electrons in the 1D channel. The $1/x$ tail 
of the potential leads to a $k\to 0$ divergence
of the Fourier transform,
\begin{eqnarray*} 
U_k &=& e^2 \int_{-\infty}^\infty dx \frac{\cos(kx)}{\kappa \sqrt{d^2+x^2}} \\
&=& (2e^2/\kappa)\, K_0(kd) \simeq (2e^2/\kappa)\,
 |\ln|kd|| \;,\qquad |kd|\ll 1\;,
\end{eqnarray*}
where $d$ is the width of the 1D channel $(k_F d\ll 1)$, and $\kappa$
denotes the dielectric constant. The Bessel function $K_0(z)$ can be
approximated by a logarithm here, and
the corresponding logarithmic correction in the plasmon dispersion 
relation (\ref{plasmon}) is
\[
\omega_k = \alpha_c v_F |k| \sqrt{|\ln|kd||} \;,
\]
where the dimensionless Coulomb interaction constant 
\[
\alpha_c =  \sqrt{2e^2/\pi\kappa v_F}
\]
is of order unity in typical quantum wires \cite{gogolinrev}.

\begin{figure}
\epsfysize=8cm
\epsffile{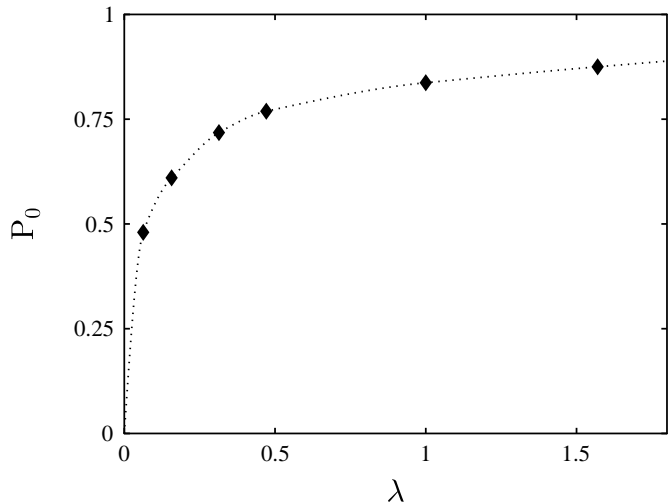}
\caption[]{\label{fig6} Monte Carlo data for $P_0$ as
a function of impurity strength in the long--ranged case.
 We have taken $\alpha_c=2.5$ and $d=\pi v_F/\omega_c$.}
\end{figure}

Unfortunately, the boson propagator (\ref{bosprop})  cannot be 
evaluated in closed form anymore. However, one can obtain the asymptotic
expansion for large $x$, 
\begin{equation} \label{flr}
F(x,\omega) = \frac{2v_F}{x\omega^2} \sin\left[\frac{|\omega| x}
{\alpha_c v_F \sqrt{|\ln|\omega||}}\right]\;.
\end{equation}
Remarkably, up to a prefactor of order unity, this expression holds 
even at $x=0$, such that \cite{fabrizio}
\[
F(0,\omega) \sim \frac{1}{|\omega| \sqrt{|\ln|\omega||} }\;.
\]
Computing the pinning function by Monte Carlo, we find that 
$P(x)$ is essentially constant, $P(x) \simeq P_0$,
as could have been anticipated from the corresponding limit $g\to 0$ of the 
Luttinger liquid case.  The functional dependence of $P_0$ on the impurity 
strength is depicted in Fig.~\ref{fig6}.  

For long--ranged interactions,
the decay of the Friedel oscillation away from the barrier is 
completely governed by the envelope function $W(x)$ specified in 
Eq.~(\ref{envel}). Asymptotic evaluation of the frequency integral
using (\ref{flr}) shows that for $x\gg d$, the leading contribution is
\[
W(x) \sim \exp \left[ - {\rm const.}\sqrt{\ln(x/d)} \right ]\;.
\]
This decay is slower than any power law.
In conclusion,  in the presence of long--ranged unscreened 
Coulomb interactions, the Friedel oscillation takes the following
form far away from the barrier
\begin{equation} \label{wlr}
\delta \rho(x) \sim \cos(2k_F x)  \, 
\exp \left[ -{\rm const.} \sqrt{\ln x} \right ]\;.
\end{equation}
This long--ranged density
disturbance should lead to strong quasi--Bragg peaks  in x--ray
scattering.

\subsection{Spin--$\frac12$ case}
 
Let us now briefly comment on the spin--$\frac12$ case. The 
bosonized treatment proceeds very similarly \cite{prl}. One can introduce
charge fields $\theta_\rho(x), \phi_\rho(x)$ generalizing the fields 
$\theta(x)$ and $\phi(x)$ employed 
previously,  and, in addition, spin fields
 $\theta_\sigma(x)$ and $\phi_\sigma(x)$. These fields are 
linear combinations of the respective spin--up and --down fields, such that
the electron creation operator for spin $s=\pm$ at location $x$ 
takes the form
\begin{eqnarray*}
\psi_s^\dagger(x) &=& \sqrt{\frac{\omega_c}{2\pi v_F}} \; \sum_{p=\pm}
\exp \left[ip\left(k_F x+\sqrt{\pi/2} \,[\theta_\rho(x)+
s\theta_\sigma(x)]\right)
\right]\\&\times& \exp\left[ 
 i \sqrt{\pi/2} \,[\phi_\rho(x)+s \phi_\sigma(x) ] \right]\;.
\end{eqnarray*}
The various $\theta_\nu$ fields ($\nu=\rho,\sigma$) commute among themselves, 
as do the $\phi_\nu$ fields. The correct generalization of (\ref{alg})
is then given by
\[
[ \phi_\nu(x),\theta_{\nu'}(x')]_- =  -(i/2) \delta_{\nu \nu' }
 \,\mbox{sgn}(x-x')\;,
\]
and the canonical momentum  for the $\theta_\nu$ field is 
$\Pi_\nu  = \partial_x\phi_\nu$.  

We are concerned with density distributions in the presence of impurities or
barriers. Generalizing Eq. (\ref{bosondens}), the 
bosonized form of the density operator for spin--$\frac12$
electrons is 
 \begin{eqnarray*}
\rho(x) &=& \rho_0 + \sqrt{2/\pi}\, \partial_x\theta_\rho(x)
+ \frac{2k_F}{\pi} \, \cos[2k_F x+\sqrt{2\pi}\,\theta_\rho(x)]
\cos[\sqrt{2\pi}\,\theta_\sigma(x)]\\
&+&  {\rm const.} \cos[4k_F x + \sqrt{8\pi}\, \theta_\rho(x)]\;,
\end{eqnarray*}
where the background charge is now $\rho_0=2k_F/\pi$.
The three other terms 
are (1) the long--wavelength contribution, (2) the $2k_F$ charge 
density wave part, and (3) the $4k_F$ 
Wigner component \cite{emery,wigner}. The Wigner component is not present 
in the spinless case, since two right--movers have to be
flipped into left--movers simultaneously for this term to arise.
Due to the Pauli principle, 
this can only happen for spin--$\frac12$ electrons.

Assuming to be away from lattice or spin density wave
instabilities, and neglecting electron--electron 
backscattering for the moment,
the clean system is described by $H_0  = H_\sigma + H_\rho$ with
\[
H_\rho= \frac{v_F}{2} \int dx 
\left[\Pi_\rho^2(x) + (\partial_x \theta_\rho(x))^2\right]
 + \frac{1}{\pi} \int dx dx' \,\partial_x\theta_\rho(x) U(x-x') 
\partial_{x'} \theta_\rho(x')\;.
\]
The spin part $H_\sigma$ is identical to the charge part $H_\rho$
with no interaction potential $U$ and the $\rho$ fields replaced 
by the $\sigma$ fields. Furthermore, the impurity backscattering contribution
has the form
\[
H_I= V  \cos[\sqrt{2\pi}\, \theta_\rho(0)] \, 
\cos[\sqrt{2\pi}\, \theta_\sigma(0)]\;.
\]
Spin and charge parts are now coupled through this term. 
The forward-scattering contribution $\sim \partial_x \theta_\rho(0)$
leads only to a phase shift and is readily included by a gauge 
transformation.

Friedel oscillations can again be extracted from a generating functional
generalizing Eq.~(\ref{gener}),
\[
Z(x,\mu^{}_\nu) = \left\langle  \exp
\left[\sqrt{2\pi} \, i \sum_{\nu = \rho,\sigma} 
\mu^{}_\nu \theta^{}_\nu(x)  
\right] \right \rangle\;.\nonumber
\]
Since the impurity influences $\theta_\nu$ only at $x=0$,
we constrain the field amplitudes $\theta_\nu(x=0)$ to be equal to new fields,
$q_\nu(\tau) =\sqrt{2\pi}\, \theta_\nu(0,\tau)$. We then proceed as before,
and in the end,
one is left with the nontrivial average over the $q_\nu$ fields
alone, which are coupled to each other through $H_I$.

Collecting together all terms, we obtain
\begin{equation}\label{zres}
Z = - {\cal P}(x) 
\prod_{\nu=\rho,\sigma} W^{\mu_\nu^2/2}_\nu(x)
\;.
\end{equation}
The envelope functions $W_\nu(x)$ generalizing (\ref{envel}) are 
\[
W_\nu(x) = \exp\left( \int\frac{d\omega}{2\pi} 
\frac{F^{(\nu )\,2}(x,\omega)-F^{(\nu )\,2}(0,\omega)}{ F^{(\nu )}(0,\omega)}
\right)\;.
\]
The charge and spin boson propagators are defined as in Eq.~(\ref{bosprop}),
and for a Luttinger liquid, we have again a single interaction constant 
$g=g_\rho\leq 1$. In the absence of a magnetic field and any spin-dependent
interactions, one has to put $g_\sigma=1$ to respect the SU(2) spin
symmetry \cite{haldane,kane92}. Thus, the boson propagator functions read
\[
F^{(\nu)}(x,\omega) = \frac{\pi g_\nu}{|\omega|}
 \exp[-|g_\nu \omega x|/v_F] \;.
\]
Finally, the  quantity ${\cal P}$ in Eq. (\ref{zres}) generalizes the
pinning function. We find
\[
{\cal P}(x) =  - \Biggl\langle \prod_{\nu=\rho,\sigma}
 \exp\Biggl( 
i\mu_\nu \left[ q_{\nu,0} + \int \frac{d\omega}{2\omega} 
 q_\nu(\omega) \frac{F^{(\nu)}(x,\omega)}{F^{(\nu)}(0,\omega)}
\right ] \Biggr) \Biggr\rangle_q\;. 
\]
The $q$ bracket stands for an average taking the action
\[
S[q_\nu] = \sum_{\nu=\rho,\sigma} \int \frac{d\omega}{2\pi}
\frac{|\omega|}{2\pi g_\nu} |q_\nu(\omega)|^2 
+ V\int d\tau \cos [q_{\sigma,0}+ q_\sigma(\tau)]
\cos [q_{\rho,0}+ q_\rho(\tau)] \;.
\]
In the following, we only discuss asymptotic properties
far away from the barrier. For these, we can assume that ${\cal P}$
is constant such that only the envelope functions $W_\nu(x)$ have
to be evaluated. The spin part is like in the noninteracting case
and gives rise to a $1/\sqrt{x}$ factor in the 
$2k_F$ Friedel oscillation. We note that there is the same
crossover as discussed above, with a slower decay
of the Friedel oscillation at $x\ll x_0$.

Combined with the charge channel, we find an asymptotic
$\sim x^{-(1+g)/2}$ decay for the $2k_F$ Friedel oscillation.
This is slightly faster than the corresponding $x^{-g}$ law
for spinless electrons and reflects that one starts to go away 
from the extreme 1D case by incorporating the second (spin) channel. 
As electron--electron backscattering cannot be treated by a simple
renormalization of $g$ in the spin--$\frac12$ case, there could be
weak logarithmic corrections to the $\sim x^{-(1+g)/2}$ law,
depending on the magnitude of the backscattering amplitude \cite{solyom}.
Generalizing the two--channel spin--degenerate case discussed in this
subsection to a multi--channel situation, one would then
expect a $x^{-1}$ law for large channel number,
since Fermi liquid theory will eventually be valid
in that case \cite{glaz2}.

Remarkably, for spin--$\frac12$ electrons,
there is also a $4 k_F$ Friedel oscillation component
\[
 \delta \rho (x)  \sim \cos(4k_{\rm F}x)\,x^{-2g} \;, 
\]
which dominates over the $2 k_F$ contribution for  
strong enough correlations, $g<\frac13$. Since $4 k_F$ corresponds 
to the inter--particle spacing, this suggests that for $g<\frac13$
signatures of Wigner crystal behavior are induced by the
impurity.  The $4k_F$ component has been seen
in numerically exact small--chain calculations by Hallberg
and Balseiro \cite{hallberg}. In their Lanczos calculation,
for a Hubbard chain with a magnetic impurity,
the $4k_F$ component was found to be present for strong
interactions.

Wigner crystal behavior has also been found by Schulz \cite{wigner}
for the clean system with long--ranged correlations.
For $1/x$ interactions, the $4 k_F$ Friedel 
oscillation decay is extremely slow. While the spin  
degrees of freedom involve again the $1/\sqrt{x}$ factor
suppressing the $2k_F$ component, the  
$4 k_F$ Friedel oscillations decay again like $\exp(-c \sqrt{\ln x})$,
i.e., slower than any power law.
Effectively, one will then only observe the $4k_F$ 
component. 
In the spinless case,  the same quasi long--ranged 
behavior appears for the $2 k_F$ component already because the
spin channel is absent, see Eq.~(\ref{wlr}).

\subsection{Friedel oscillations near a double barrier}
\label{sec:dble}

Let us now consider a double barrier arrangement which allows
for the possibility of perfect transmission even in presence
of Coulomb interactions. The resonant tunneling problem for
a Luttinger liquid has been studied extensively before \cite{rt1},
and here we shall focus on the Friedel oscillation for a spinless
single--channel Luttinger liquid. 
The symmetric double barrier structure considered is described
by the Hamiltonian
\begin{equation} \label{impurity2}
H_I = V \sum_{\sigma=\pm} \cos[2\sqrt{\pi}\theta(\sigma R/2)] 
\end{equation}
instead of the single--barrier expression (\ref{impurity}). 
We omit the forward-scattering contribution since it is inessential
for the subsequent discussion.  The 
resonance can then be tuned by varying $k_F R$ (or by varying a
gate voltage coupling to the charge on the ``island'' formed between
the barriers), and we employ again a dimensionless impurity strength
$\lambda=\pi V/\omega_c$.

To study the Friedel oscillation  for the interacting double--barrier
problem, we proceed as in the single--barrier case and compute the
generating functional (\ref{gener}).
The $\theta$ field can be integrated out by introducing the two fields
\[
q_\pm = 2\sqrt{\pi} \theta(\pm R/2) \;,
\]
where these constraints are enforced by Lagrange multiplicator 
fields $\Lambda_\pm(\tau)$. The Gaussian integration over the
$\theta$ degrees of freedom and the ensuing minimization of the Lagrange 
multiplier fields leave us with the following expression for
the Friedel oscillation
\[
\delta \rho(x) /\rho_0 = 
- W_2(x) \,{\rm Re}\, e^{2 i k_F |x|} \, P_2(x)
\]
with $\rho_0=k_F/\pi$. The ground--state envelope function 
takes the general form 
\begin{eqnarray*} 
W_2(x) &=& \exp \int \frac{d\omega}{2\pi}  \Biggm\{ -F(0,\omega)+
 \frac{F(0,\omega)}
{F^2(0,\omega) - F^2(R,\omega)} \\ 
&\times& \biggm[ F^2(|x+R/2|,\omega) +
F^2(|x-R/2|,\omega) \\ &-& 2 \frac{F(R,\omega)}{F(0,\omega)}
F(|x+R/2|,\omega) F(|x-R/2|,\omega) \biggm] \Biggm\} \;,
\end{eqnarray*}
with the boson propagator $F(x,\omega)$ defined
in Eq.~(\ref{bosprop}).
For the spinless Luttinger liquid, this gives to the right of
the right barrier 
\[
W_2(x) = [1+(x-R/2)/\alpha]^{-g} \;,\qquad x>R/2\;,
\]
where again $\alpha=v_s/2\omega_c$.

The pinning function $P_2(x)$ contains all dependency on impurity 
properties and can be expressed as average in $\{ q_+, q_- \}$
space. It is more  convenient to switch to the symmetric and
antisymmetric combinations \cite{rt1}
\begin{eqnarray*}
q &=& (q_+ + q_-)/2 \\
Q &=& (q_+ - q_-)/2 
\end{eqnarray*}
describing the transmitted charge $q$ and the island charge 
$Q$, respectively. The action for the impurity averaging 
reads for finite temperatures
\begin{eqnarray}\nonumber
S[q,Q]&=&  
\frac{1}{\beta} \sum_{n=-\infty}^\infty
\frac{|\omega_n|}{2\pi g} 
\frac{q_n q_{-n}}{1+\exp(-|\omega_n| R / v_s)} \\ &+& \nonumber 
\frac{1}{\beta} \sum_{n=-\infty}^\infty
\frac{|\omega_n|}{2\pi g} 
\frac{Q_n Q_{-n}}{1-\exp(-|\omega_n| R / v_s)}  \\
&+& 2V \int_0^\beta d\tau \,\cos[q(\tau)]\,
\cos[k_F R+Q(\tau)] \;,\label{actrt}
\end{eqnarray}
where $q_n$ and $Q_n$ are Matsubara components
at frequency $\omega_n=2\pi n/\beta$.
The two modes are  coupled by the impurity term.
The $Q$ mode has acquired a mass gap now \cite{kane92}.

For the spinless Luttinger liquid, the
 complex--valued pinning function $P_2(x)$ takes the general form
\begin{eqnarray*}
P_2(x) & =& -\Bigl\langle \exp\Bigl[\frac{i}{\beta}\sum_n 
\Bigl( 
\frac{e^{-|\omega_n (x+R/2)|/v_s} + e^{-|\omega_n (x-R/2)|/v_s}}
{1 + e^{-|\omega_n| R/v_s}} \;q_n
 \\ \nonumber &+& 
\frac{e^{-|\omega_n (x-R/2)|/v_s} - e^{-|\omega_n (x+R/2)|/v_s}}
{1 - e^{-|\omega_n| R/v_s}}\;
 Q_n \Bigr) \Bigr] \Bigr\rangle_{q,Q} \;.
\end{eqnarray*}
To the right of the right barrier, $x>R/2$, this yields the simpler 
expression
\[
P_2(x) = -\left\langle \exp\left[\frac{1}{\beta}
\sum_n \, e^{-|\omega_n|(x-R/2)/v_s} (Q_n + q_n) \right] \right\rangle_{q,Q}
\]
where the average has to be taken using the action (\ref{actrt}).

In the following, we shall focus on the case of small barriers.
This has two reasons. (1) In the strong--scattering limit the 
resonances are very sharp, and one might not be able to ``find''
the on--resonance Friedel oscillation. (2) A
strong--scattering SCHA treatment, as applied to the single--barrier case 
previously, is not possible here since SCHA neglects 
tunneling and hence does not capture resonant tunneling. The SCHA
prediction is always $x^{-g}$, which
is only correct in the limit $\lambda\to \infty$
where the barriers reflect completely and no resonance is possible anymore.

In the weak--scattering case, we may compute the pinning function
 perturbatively in the impurity strength.
To lowest order in $\lambda$, we obtain the Friedel oscillation 
\begin{equation} 
\delta \rho(x)/\rho_0  = - \lambda \gamma_g^{(1)}
 \sum_{\sigma=\pm} \cos\Big(2k_F(x-\sigma R/2)\Big) \
 [(x-\sigma R/2)/\alpha]^{1-2g}  \label{pertrt}
\;,
\end{equation}
with $\gamma_g^{(1)}$ defined in Eq.~(\ref{ggg}).
This result is only valid for $g>\frac12$ and $\lambda\ll 1$. Off resonance,
we find the usual $x^{1-2g}$ dependence, see Eq.~(\ref{pertu}). 
On resonance, the $2k_F$ component of the Fourier transform 
$V_k= 2V \cos(k R/2)$ of the scattering potential (\ref{impurity2}) 
vanishes \cite{kane92}. Hence, the two cosine terms in Eq. (\ref{pertrt})
have different signs and 
interfere destructively. In that case, the Friedel oscillation
has a different  asymptotic behavior at $x\gg R$, 
\begin{equation} \label{x2g}
\delta \rho(x) /\rho_0= -\lambda \,(2g-1) \gamma_g^{(1)}
\cos[2k_F(x-R/2)]\, \frac{R}{\alpha} \left(\frac{x}{\alpha}\right)^{-2g} \;.
\end{equation}
The third--order contribution $\sim \lambda^3$ goes like $x^{-4g}$ 
on resonance. Therefore, Eq.~(\ref{x2g}) holds in the asymptotic
regime, contrary to the off--resonance case, where the higher order terms
decay slower than the lowest--order contribution and thus lead to a 
modification of the perturbative prediction in the asymptotic 
regime (instead of $x^{1-2g}$, the correct asymptotic off--resonance
law is $x^{-g}$).

In conclusion, for resonant tunneling through a double barrier, the Friedel 
oscillation can be dramatically affected if the barrier is tuned on resonance.
The usual $x^{-g}$ law is turned into the faster $x^{-2g}$ law
far away from the scatterers. This should be observable as a much
weaker signal in x--ray scattering or NMR data.

\section{CONCLUDING REMARKS}

In this article, we have applied
the bosonization method to compute Friedel oscillations
for interacting fermions in one spatial dimension. The success
of this method might come as a surprise, since density properties
usually depend on details of the (noninteracting)
electron band, such as band curvature. However, the asymptotic
properties (which means several lattice spacings
away from the barrier in that context) depend only on Fermi
surface quantities and can be computed by bosonization.

We have shown that in 1D the Friedel oscillation in a correlated
fermionic system decays slower than in the Fermi liquid case.
This is in conflict with naive
expectations based on the smeared momentum distribution of a Luttinger
liquid. While in a Fermi liquid thermal 
smearing of the Fermi distribution
induces an exponential decay of the Friedel oscillation,
the smeared momentum distribution of a Luttinger liquid does
not imply a faster decay. This shows again the failure of
the quasiparticle picture for correlated fermions in 1D.

The results provided here have intrinsic many--body character and,
to the best of our knowledge,  
cannot be obtained by other methods available at the moment. 
While results for the conductance can be obtained for weak
interactions from a Hartree--Fock
type approach \cite{matveev1}, it is not possible
to obtain the correct density profile with such methods. 
Furthermore, the fermionization
technique allowing for exact results for the special interaction
constant $g=\frac12$  fails for the density profile
as well. We hope our findings will motivate 
further theoretical work, e.g., on the effects of 
{\em magnetic} impurities on the screening cloud
in one--dimensional interacting fermions.

\acknowledgements
This article requires basic knowledge of bosonization techniques
as provided in the article by M. Sassetti in the same volume. 
The help by A. Komnik in preparing the figures is gratefully
acknowledged.

\end{document}